\documentclass[submission,copyright,creativecommons]{eptcs}
\providecommand{\event}{MCU 2013} 
\usepackage{breakurl}             
\usepackage{graphicx}

\newtheorem{lem}{Lemma}

\newtheorem{fig}{Figure}

\newtheorem{automaton}{Automaton}

\def\leurre{\noindent\leftskip0pt\small\baselineskip 10pt}

\def\encadre#1#2{%
\setbox100=\hbox{\kern#1{#2}\kern#1}
\dimen100=\ht100 \advance \dimen100 by #1
\dimen101=\dp100 \advance \dimen101 by #1
\setbox100=\hbox{\vrule height \dimen100 depth \dimen101\box100\vrule}
\setbox100=\vbox{\hrule\box100\hrule}
\advance \dimen100 by .4pt \ht100=\dimen100
\advance \dimen101 by .4pt \dp100=\dimen101
\box100
\relax
}

\def\ligne#1{\hbox to \hsize{#1}}
\def\PlacerEn#1 #2 #3 {\rlap{\kern#1\raise#2\hbox{#3}}}

\def\grostrait{\ligne{\vrule height 1pt depth 1pt width \hsize}}
\def\demitrait{\ligne{\vrule height 0.5pt depth 0.5pt width \hsize}}

\def\reunion{\mathop{\cup}}

\title{Hyperbolic tilings and formal language theory}
\author{Maurice Margenstern$^1$
\institute{Universit\'e de Lorraine,\\ LITA EA3097,\\ Campus du Saulcy,\\
57045 C\'edex, {\sc France}
}
\email{maurice.margenstern@univ-lorraine.fr}
\email{margenstern@gmail.com}
\and 
K.G. Subramanian$^2$,
\institute{$^2$ School of Computer Sciences,\\
Universiti Sains Malaysia,\\ 11800 Penang, Malaysia,
}
\email{kgsmani1948@yahoo.com}
}
\def\titlerunning{Hyperbolic tilings and formal language theory}
\def\authorrunning{M. Margenstern, K.G. Subramanian}
\begin{document}
\maketitle

\begin{abstract}
In this paper, we try to give the appropriate class of languages to which belong various
objects associated with tessellations in the hyperbolic plane. 
\end{abstract}

ACM-class: 
F.2.2., F.4.1, I.3.5

{\bf keywords}:
pushdown automata, iterated pushdown automata, tilings, hyperbolic plane, 
tessellations

\vskip 10pt

\def\cqfd{\hbox{\kern 2pt\vrule height 6pt depth 2pt width 8pt\kern 1pt}}
\def\Hii{$I\!\!H^2$}
\def\Hiii{$I\!\!H^3$}
\def\Hiv{$I\!\!H^4$}
\def\norm{\hbox{$\vert\vert$}}
\section{Introduction}

   In \cite{mmISRN}, it was shown that a few languages constructed from some figures of 
hyperbolic tilings cannot be recognized by pushdown automata but they can be recognized by 
a 2-iterated pushdown automaton.
Before, it was known that several tessellations of the hyperbolic plane are generated
by substitutions, see~\cite{gms-alg}. This property is also clear from ~\cite{mmJUCStile}.

   These substitutions can be also described by the use of grammars. This is rather 
straightforward. In \cite{mmJUCSii}, these substitutions appear as rules of a grammar,
although the grammar is not formally described.

   Iterated pushdown automata were introduced in~\cite{greibach,maslov} and we refer
the reader to~\cite{fratani-senizergues} for references and for the connection of this
topic with sequences of rational numbers. By their definition, iterated pushdown 
automata are more powerful than standard pushdown automata but they are far less powerful
than Turing machines. As Turing machines can be simulated by a finite automaton with
two independent stacks, iterated pushdown automata can be viewed as an intermediate
device, see also \cite{kelarev} for other connections of automata with graph
algebras.

   In this paper, we show an application of this device to the characterization
of contour words of a family of bounded domains in many tilings of the hyperbolic
plane. We can do the same kind of application for a tiling of the hyperbolic $3D$ space
and for another one in the hyperbolic $4D$ space. These two latter applications cannot
be generalized to any dimension as, starting from dimension~5, there is no tiling of the
hyperbolic space which would be a tessellation generated by a regular polytope.

   In Section~\ref{pushdown}, we remember the definition of iterated pushdown automata with
an application to the computation of the recognition of words of the form $a^{f_n}$, where
$\{f_n\}_{n\in I\!\!N}$ is the Fibonacci sequence with $f_0=f_1=1$. This sequence
will always be denoted by $\{f_n\}_{n\in I\!\!N}$ throughout the paper.

   In Section~\ref{htilings}, we remind the reader about several features and properties 
on tilings of the hyperbolic plane.

   In Section~\ref{grammars}, we indicate how several tilings can be defined by
a grammar.

   In Section~\ref{wcontours}, we define the contour words which we are 
interested in and we
construct iterated pushdown automata which recognize them for the case
of the pentagrid and the heptagrid, {\it i.e.} the tilings $\{5,4\}$ and $\{7,3\}$ 
of the hyperbolic plane. 
In the same section, we extend these results to infinitely 
many tilings of the hyperbolic plane.


\section{Iterated pushdown automata}
\label{pushdown}

   In this section, we fix the notations which will be used in the paper. We follow
the notations of \cite{fratani-senizergues}.

\subsection{Iterated pushdown stores}

   This data structure is defined by induction, as follows:

   {\leftskip 20pt\parindent 0pt
   0-pds($\Gamma$) = $\{\epsilon\}$\vskip 2pt
   $k$+1-pds($\Gamma$) = $(\Gamma[k$-pds$(\Gamma)])^*$\vskip 2pt
   it-pds($\Gamma$) = $\reunion_{k}k$-pds$(\Gamma)$
   \par}

   The elements of a $k$+1-pds($\Gamma$) structure are $k$-pds$(\Gamma)$ structures
and each element is labelled by a letter of~$\Gamma$. A $k$-pds$(\Gamma)$ structure
will often be called a $k$-level store, for short. When $k$ is fixed, we speak of
{\it outer} stores and of {\it inner} stores in a relative way: an $i$-level store
is {\bf outer} than a $j$-one if and only if $i< j$. In the same situation, the
$j$-level store is {\bf inner} than the $i$-one.

   We define functions and operations on $k$-level stores, by induction 
on~$k$.

   From the above definition, we get that a $k$+1-level store~$\omega$ can be uniquely 
represented in the form:

\ligne{\hfill $\omega = A[flag].rest$,\hfill}

\noindent
where $A\in\Gamma$, $flag$ is a $k$-level store and $rest$ is $k$+1-store. Moreover,
if $\ell$ is the number of elements of~$rest$, the number of elements of~$\omega$
is $\ell$+1.

   A first operation consists in defining the generalization of the standard notion of 
{\it top symbol} in an ordinary pushdown structure. This is performed by the function
$topsym$ defined by:

   {\leftskip 20pt\parindent 0pt
   $topsym(\epsilon) = \epsilon$\vskip 2pt
   $topsym(A[flag].rest) = A.topsym(flag)$
   \par}

It is important to remark that $topsym$ is the single direct access to all inner stores
of a $k$-level store. In other words, for any inner store, only its topmost symbol can
be accessed and when this inner store is in the top of the outmost store. 

   Also note that the $topsym$ function performs a {\bf reading}. There are two 
families of {\bf writing} operations, also concerning the elements visible from 
the $topmost$ function only.

   The first one consists of the $pop$ operations defined by the following induction:

   {\leftskip 20pt\parindent 0pt
   $pop_j(\epsilon)$ is undefined \vskip 2pt
   $pop_{j+1}(A[flag].rest) = A.[pop_j(flag)].rest$
   \par}

   The second family consists of the $push$ operations defined by the following induction:

   {\leftskip 20pt\parindent 0pt
   $push_1(\gamma)(\epsilon) = \gamma$, for $\gamma\in\Gamma$ \vskip 2pt
   $push_j(\gamma)(\epsilon)$ is undefined for $j> 1$ \vskip 2pt
   $push_{j+1}(w)(A[flag].rest) = w_1[flag]..w_k[flag].rest$,
where $w = w_1..w_k$, with $w_i\in \Gamma$ for $1\le i\le k$
   \par}

\subsection{Iterated pushdown automata}

   Intuitively, the definition is very close to the traditional one of standard
non-deterministic standard push-down automata. A $k$-iterated pushdown automaton is 
defined by giving the following data:

{\leftskip 20pt\parindent 0pt
- a finite set of states, $Q$;\vskip 2pt
- an input finite alphabet $\Sigma$;\vskip 2pt
- a store finite alphabet $\Gamma$;\vskip 2pt
- a transition function $\delta$ from $Q\times\Sigma\cup\{\epsilon\}\times\Gamma^k$ into
a finite set of instructions of the form $(q,${\bf op}$)$, where $q$~is a state
and {\bf op} is a pop- or a push-operation as described in the previous sub-section. 
\par}
  
   We also assume that there is an initial state denoted by~$q_0$ and that the initial
state of the store is $Z[\epsilon]$, where $Z$~is a fixed in advance symbol of~$\Gamma$.
Note that we allow $\epsilon$-transition which play a key role.

   A configuration is a word of the form $(q,w,\omega)$, where $q$~is the current state 
of the automaton, $w$ is the current word and $\omega$ is the current $k$-level store of the 
automaton. A computational step of the automaton allows to go from one configuration
to another by the application of one transition. In order to apply a transition, 
the current state 
of the automaton must be that of the transition, the first letter of~$w$ must 
be the symbol
of~$\Sigma$ in the transition if any, and $topsym(\omega)$ must be the word of
$\Gamma^k$ in the transition if any. A word $w$ is accepted if and only there
is a sequence of computational steps starting from $(q_0,w,Z[\epsilon])$ to
a first configuration of the form $(q,\epsilon,\epsilon)$. The language recognized
by a $k$-iterated pushdown automaton is the set of words in $\Sigma^*$ which are
accepted by the automaton.

\subsection{An example: the Fibonacci sequence}
   
  As an illustrative example of the working of such an automaton, we take the
set of words of the form $a^{f_n}$, where  $\{f_n\}_{n\in I\!\!N}$ is the
Fibonacci sequence. This language is recognized by a 2-iterated pushdown automaton
as proved in~\cite{fratani-senizergues}. Here, we give the automaton and a proof of
its correctness.

\vtop{
\begin{automaton}\label{autom1}
\leurre
The $2$-pushdown automaton recognizing the Fibonacci sequence.
\end{automaton}
\vspace{-12pt}
\grostrait
\vskip 0pt
three states: $q_0$, $q_1$ and~$q_2$; input word in $\{a\}^*$;
$\Gamma = \{Z,X_1,X_2,F\}$;
\vskip 0pt\noindent
initial state: $q_0$; initial stack: $Z[\epsilon]$; transition function $\delta$:
\vskip 0pt
\vspace{-4pt}
\demitrait
\vskip 0pt
{\leftskip 20pt\parindent 0pt
$\delta(q_0,\epsilon,Z) = \{(q_0,push_2(F)),(q_0,push_1(X_2))\}$\vskip 2pt
$\delta(q_0,\epsilon,ZF) = \{(q_0,push_2(FF)),(q_0,push_1(X_2))\}$\vskip 2pt
$\delta(q_0,\epsilon,X_1F) = (q_1,pop_2)$\vskip 2pt
$\delta(q_0,\epsilon,X_2F) = (q_2,pop_2)$\vskip 2pt
$\delta(q_0,a,X_1) = (q_0,pop_1)$\vskip 2pt
$\delta(q_0,a,X_2) = (q_0,pop_1)$\vskip 2pt
$\delta(q_1,\epsilon,X_1F) = (q_0,push_1(X_1X_2))$\vskip 2pt
$\delta(q_2,\epsilon,X_2F) = (q_0,push_1(X_1))$\vskip 2pt
$\delta(q_1,\epsilon,X_1) = (q_0,push_1(X_1X_2))$\vskip 2pt
$\delta(q_2,\epsilon,X_2) = (q_0,push_1(X_1))$\vskip 2pt
\par}
\vskip 0pt
\demitrait
}
\vskip 10pt
The proof is based on the following lemma:

\begin{lem}\label{lem1}
We have the following relations, for any non-negative~$k$:
\vskip 5pt
   $(q_0,a^{f_k},X_2[F^k].\omega) \Rightarrow_\delta^* (q_0,\epsilon,\omega)$\vskip 2pt
   $(q_0,a^{f_{k+1}},X_1[F^k].\omega) \Rightarrow_\delta^* (q_0,\epsilon,\omega)$
\end{lem}

\noindent
Proof. It is performed by induction whose basic case $k=0$ is easy.
If we start from $(q_0,a^{f_{k+1}},X_1[F^k].\omega)$, we have the following
derivation:

{\leftskip 20pt\parindent 0pt
 $(q_0,a^{f_{k+2}},X_1[F^{k+1}].\omega) \vdash  (q_1,a^{f_{k+2}},X_1[F^k].\omega)$\vskip 2pt
 $\vdash  (q_0,a^{f_{k+2}},X_1[F^k].X_2[F^k].\omega)
 \vdash (q_0,a^{f_k},X_2[F^k].\omega)$
\par}

\noindent
by induction hypothesis as $f_{k+2} = f_{k+1}+f_k$. And, again by induction hypothesis:

{\leftskip 20pt\parindent 0pt
 $(q_0,a^{f_k},X_2[F^k].\omega) \vdash (q_0\epsilon,\omega)$
\par}

   Similarly,

{\leftskip 20pt\parindent 0pt
 $(q_0,a^{f_{k+1}},X_2[F^{k+1}].\omega) \vdash  (q_2,a^{f_{k+1}},X_2[F^k].\omega)
 \vdash (q_0,a^{f_{k+1}},X_1[F^k].\omega)$\vskip 2pt
 $\vdash  (q_0,\epsilon,\omega)$,
\par}

\noindent
by induction hypothesis.

   Let $a^m$ be the initial word. With the first two transitions, we guess an integer~$k$
such that $m = f_k$ if any. Then we arrive to the configuration $(q_0,a^m,Z[F^k])$.
Next, we have:

{\leftskip 20pt\parindent 0pt
$(q_0,a^m,Z[F^k]) \vdash (q_0,a^m,X_2[F^k])$.
\par}

\noindent
And by the lemma, we proved that $(q_0,a^m,X_2[F^k])\vdash (q_0,\epsilon,\epsilon)$ and
so, the word is accepted.

   We can see that if $m=f_k$ and if we guessed a wrong~$k$, then either the word is
not empty when the store vanishes, and we cannot restore it, or the word is empty as the
store is not. This also shows that if $m\not= f_k$, as there is in this case a unique
$k$ such that $f_k<m<f_{k+1}$, we always have either an empty word and a non-empty store
or an empty store with a non-empty word, whatever the guess.\cqfd

   Now, the motivation for taking iterated pushdown automata to recognize this language
is that the language cannot be recognized by ordinary pushdwon automata, whether deterministic
or non-deterministic. This can be proved by a simple application of Ogden's pumping lemma.
As the length of the words of the language has an exponential increasing, it cannot contain
words with a linear increasing.

\section{The tilings of the hyperbolic plane}
\label{htilings}

   We assume that the reader is a bit familiar with hyperbolic geometry, at least
with its most popular models, the Poincar\'es's half-plane and disc.

   We remember the reader that in the hyperbolic plane, thanks to a well known theorem
of Poincar\'e, there are infinitely many tilings which are generated by 
\ligne{\hfill}

\vtop{
\ligne{\hfill
\includegraphics[scale=0.6]{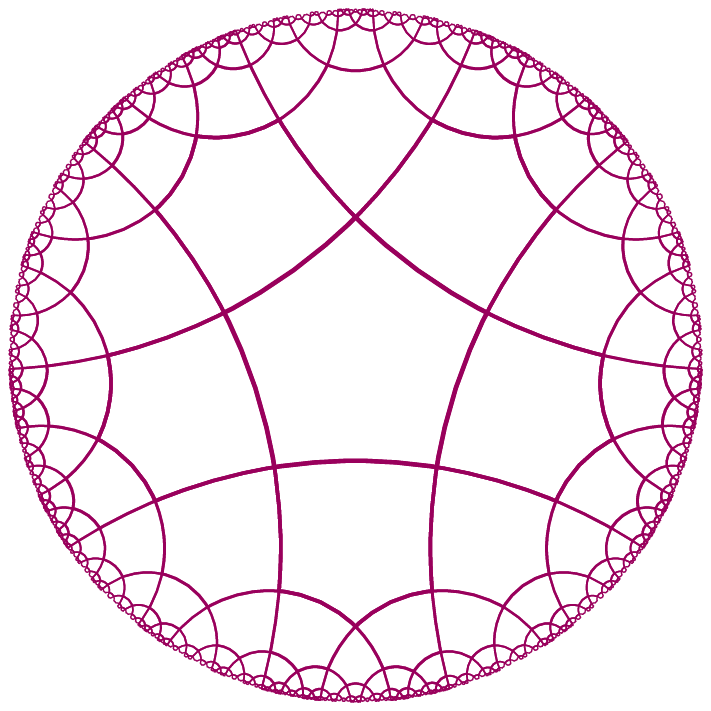}
\includegraphics[scale=0.6]{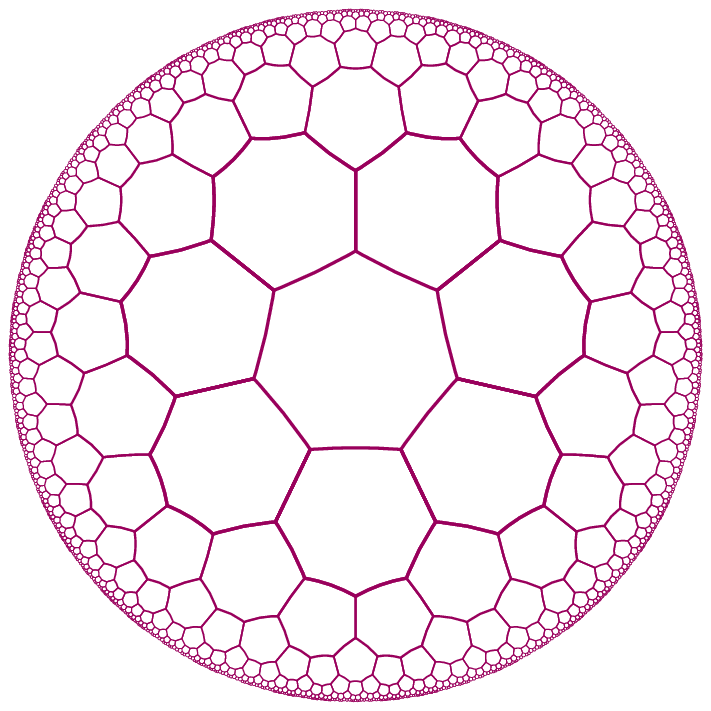}
\hfill
}
\begin{fig}\label{tilings}
\leurre
Left-hand side: the pentagrid. Right-hand side: the heptagrid. 
\end{fig}
}

\noindent
tessellation
starting from a regular polygon. This means that, starting from the polygon,
we recursively copy it by reflections in its sides and of the images in their sides.
This family of tilings is defined by two parameters: $p$, the number of sides 
of the polygon and $q$, the number of polygons which can be put around a vertex without
overlapping and covering any small enough neighbourhood of the vertex.

   In order to represent the tilings which we shall consider and the regions whose
contour word will be under study, we shall make use of the Poincar\'e's disc model.
Our illustrations will take place in the {\bf pentagrid} and the {\bf heptagrid},
{\it i.e.} the tilings $\{5,4\}$ and $\{7,3\}$ respectively of the hyperbolic plane.
Figures~\ref{tilings} and~\ref{splittings} illustrate these tilings.

\vtop{
\vskip 10pt
\ligne{\hfill
\includegraphics[scale=0.4]{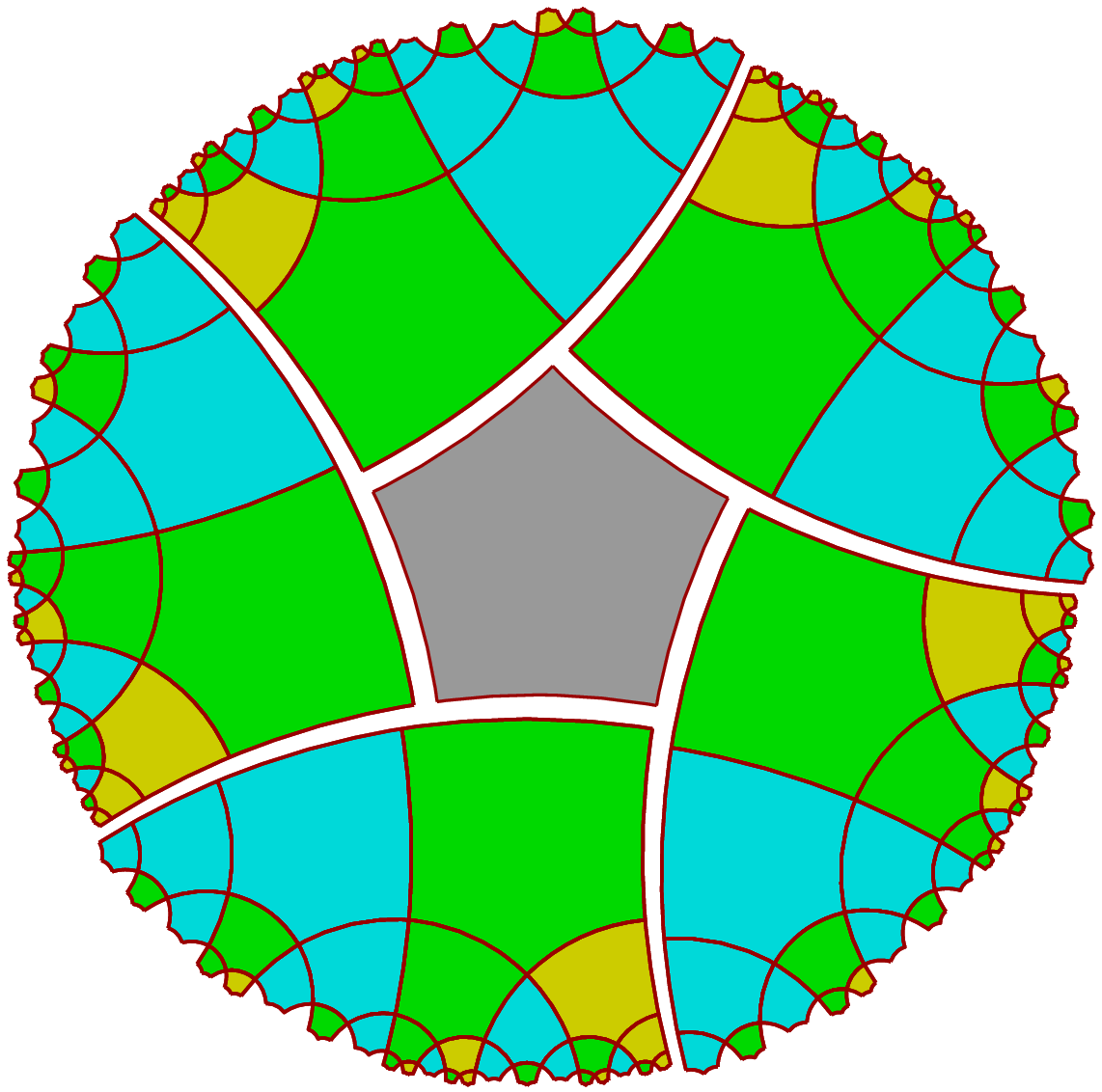}
\includegraphics[scale=0.4]{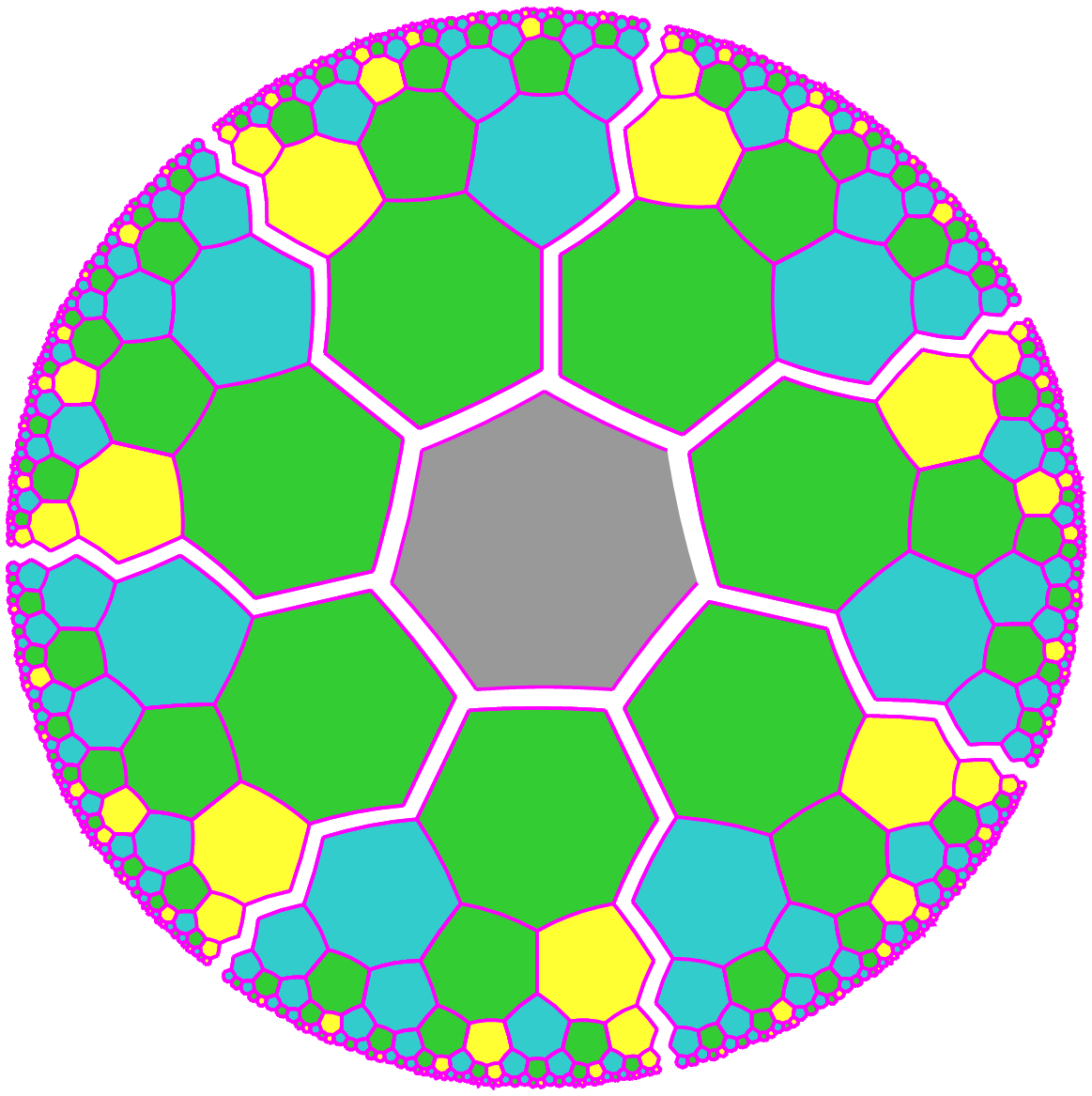}
\hfill
}
\begin{fig}\label{splittings}
\leurre
Left-hand side: the pentagrid. Right-hand side: the heptagrid. 
Note that in both cases, the sectors are spanned by the same tree.
\end{fig}
}

\vskip-5pt
\ligne{\hfill}
\vtop{
\ligne{\hfill
\includegraphics[scale=1.2]{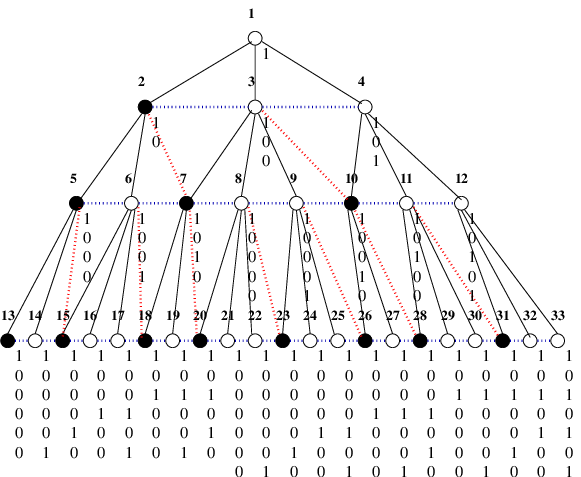}
\hfill
}
\begin{fig}\label{fibotree}
\leurre
The standard Fibonacci tree. The nodes are numbered from the root, from left to right on
each level and level after level. For each node, the figure displays the representation
of the number of the node with respect to the Fibonacci sequence, the representation
avoiding consecutive $1$'s.
\end{fig}
}

   From Figure~\ref{tilings},the pentagrid and the heptagrid seem rather different.
However, there is a tight connection between these tilings which can be seen from
Figure~\ref{splittings}. In both pictures of the latter figure, we represent
the tiling by selecting a central tile and then, by displaying as many sectors as
the number of sides
of the central tile. In each case, these sectors do not overlap 
and their union together with the central cell gives the tiling of the whole hyperbolic
plane. Now, there is a deeper common point: in both cases, each sector is spanned by
a tree which we call a Fibonacci tree for a reason which will soon be explained.
 
   As proved in \cite{mmJUCSii,mmbook1}, the corresponding tree can be defined
as follows. We distinguish two kinds of nodes, say black nodes, labelled by~$B$, and
white nodes, labelled by~$W$. Now, we get the sons of a node by the following rules:
$B\rightarrow BW$ and $W\rightarrow BWW$, the root of the tree being a white node,
see Figure~\ref{fibotree}. 
It is not difficult to see that if the root is on level~0 of the tree, the number
of nodes on the level~$k$ of the tree is $f_{2k+1}$, where $\{f_k\}_{k\in I\!\!N}$
is the Fibonacci sequence with $f_0=f_1=1$.

   The Fibonacci tree has a lot of nice properties which we cannot discuss here.
In particular, there is a way to locate the tiles of the pentagrid or the heptagrid
very easily thanks to coordinates devised from the properties of the Fibonacci tree,
see \cite{mmJUCSii,mmbook1,mmbook2}.

\section{Grammars}
\label{grammars}

   As mentioned in the introduction, the tilings considered in Section~\ref{htilings}
can be generated by a grammar.

   Consider the case of the pentagrid. Then its spanning tree can be generated by the 
following grammar:

\vskip 5pt
\ligne{\hfill$(G_0)$\hfill
$\vcenter{\vtop{\leftskip 0pt\parindent 0pt\hsize=150pt
symbols: $X$, $Y$, $Z$, $C$, $W$, $B$,\vskip 0pt
with $C$, $W$ and $B$~being terminals;\vskip 0pt
initial symbol: $Z$;\vskip 0pt
rules:\vskip 0pt
\ligne{\hskip 20pt $Z \Rightarrow CYYYYY$\hfill}
\ligne{\hskip 20pt $Y \Rightarrow WXYY$\hfill}
\ligne{\hskip 20pt $X \Rightarrow BXY$\hfill}
}}
$
\hfill}
\vskip 5pt

   Indeed, in the above rules, the symbol $\Rightarrow$ is interpreted as follows:
the tile which is on the left-hand side of the $\Rightarrow$ is replaced by the set of 
tiles which is indicated in the right-hand side of the $\Rightarrow$. In all cases, this
right-hand side set of tiles is a finite tree which consists of a root with its sons.
The root is the leftmost letter and the sons are the following letters given in the order
in the tree from left to right. 

    Note that this grammar is deterministic. Also note that the generation process may
vary: the replacement of the variables can be performed uniformly level by level, it can
be also performed following other rules. Also note that this grammar allows us to reproduce
the tree structure of the tessellation. If we wish to cover the plane only, we can simplify the
grammar to:

\vskip 5pt
\ligne{\hfill$(G_1)$\hfill
$\vcenter{\vtop{\leftskip 0pt\parindent 0pt\hsize=150pt
symbols: $X$, $Y$, $Z$, $T$,\vskip 0pt
with $T$~being terminal;\vskip 0pt
initial symbol: $Z$;\vskip 0pt
rules:\vskip 0pt
\ligne{\hskip 20pt $Z \Rightarrow TYYYYY$\hfill}
\ligne{\hskip 20pt $Y \Rightarrow TXYY$\hfill}
\ligne{\hskip 20pt $X \Rightarrow TXY$\hfill}
}}
$
\hfill}
\vskip 5pt

    In~\cite{mmJUCSii}, we considered other substitutions as, for instance, this one:

\vskip 5pt
\ligne{\hfill$(G_2)$\hfill
$\vcenter{\vtop{\leftskip 0pt\parindent 0pt\hsize=150pt
symbols: $X$, $Y$, $Z$, $C$, $W$, $B$,\vskip 0pt
with $C$, $W$ and $B$~being terminals;\vskip 0pt
initial symbol: $Z$;\vskip 0pt
rules:\vskip 0pt
\ligne{\hskip 20pt $Z \Rightarrow CYYYYY$\hfill}
\ligne{\hskip 20pt $Y \Rightarrow WYXY$\hfill}
\ligne{\hskip 20pt $X \Rightarrow BXY$\hfill}
}}
$
\hfill}
\vskip 5pt
\noindent
keeping the indication of the tree structure. Here, we obtain a different tree than the
one attached to the previous grammar. However, it spans the same tessellation if we
erase the difference between $C$, $W$ and~$B$.

    Now, in~\cite{mmJUCSii}, we proved that in fact, we have six possible set of rules,
considering that for $X$ we have two possible rules:

\ligne{\hfill $X \Rightarrow BXY$ and $X \Rightarrow BYX$ \hfill}

\noindent
and that for~$Y$, we have three of them:

\ligne{\hfill $Y \Rightarrow BXYY$, $Y \Rightarrow BYXY$ and $X \Rightarrow BYYX$ \hfill}

   We also proved that while replacing the variable by a symbol by the application of
a rule, we could switch from one set of rules to another at random: we obtain an uncountable
set of trees spanning the tessellation but we still obtain the same tessellation, once the
colours of the tiles are forgotten.

   This process can be described by a single grammar:

\vskip 5pt
\ligne{\hfill$(G_3)$\hfill
$\vcenter{\vtop{\leftskip 0pt\parindent 0pt\hsize=150pt
symbols: $X$, $Y$, $Z$, $T$,\vskip 0pt
with $C$, $W$ and $B$~being terminals;\vskip 0pt
initial symbol: $Z$;\vskip 0pt
rules:\vskip 0pt
\ligne{\hskip 20pt $Z \Rightarrow CYYYYY$\hfill}
\ligne{\hskip 20pt $Y \Rightarrow WXYY\;\vert\;WYXY\;\vert\;WYYX$ \hfill}
\ligne{\hskip 20pt $X \Rightarrow BXY\;\vert\;BYX$\hfill}
}}
$
\hfill}
\vskip 5pt
   This time, the grammar is non-deterministic and this relaxation of determinism allows us
to handle in a more synthetic expression a process which would require more elaboration
using the single notion of substitution.

    Last remark on the generation of the tessellation: using substitution or grammars,
the tessellation itself is obtained after using infinitely many applications of the rules.
Finitely many applications always lead to a finite figure whose size increases with the
number of applications.

    Also note that if we apply only rules with~$Y$, we get binary trees. If we apply
only rules with~$X$, we get six lines of~$X$ with, for each $X$-line, a kind of shadow 
consisting of~$Y$'s.  

    The grammars~$(G_1)$ up to~$(G_3)$ can be generalized to the tilings studied 
in~\cite{mmgsFI} and~\cite{mmJUCS4D},
the tessellations $\{5,3,4\}$ and $\{5,3,3,4\}$. This can also be generalized more easily
with the tilings $\{p,4\}$ and $\{p$+$2,3\}$ when $p\geq5$. For the same value of~$p$,
the tessellations $\{p,4\}$ and $\{p$+$2,3\}$ are generated by the same tree which generalizes
the Fibonacci tree. The generalizations of~$(G_1)$ are of the form:  

\vskip 5pt
\ligne{\hfill$(G_p)$\hfill
$\vcenter{\vtop{\leftskip 0pt\parindent 0pt\hsize=150pt
symbols: $X$, $Y$, $Z$, $C$, $W$, $B$,\vskip 0pt
with $C$, $W$ and $B$~being terminals;\vskip 0pt
initial symbol: $Z$;\vskip 0pt
rules:\vskip 0pt
\ligne{\hskip 20pt $Z \Rightarrow CY^p$\hfill}
\ligne{\hskip 20pt $Y \Rightarrow WXY^{p-3}$\hfill}
\ligne{\hskip 20pt $X \Rightarrow BXY^{p-4}$\hfill}
}}
$
\hfill}
\vskip 5pt
   Again, we can define \hbox{$p$$-$3} rules for~$Y$ and \hbox{$p$$-$4} rules for~$X$
and, as above, a non-deterministic grammar which can generate uncountably many trees,
each one generating the considered tessellation.

\section{Contour words and words along a level} 
\label{wcontours}
   
   In  \cite{mmISRN}, the first author considered the possibility to define
words by looking at a specific object: the set of tiles which lie at a given distance
from another tile, fixed in advance and once for all.

   Fix a tile~$C$ which will later be called the central one. A path from a tile~$T$
to~$C$ is a finite sequence $T_i$, $0\leq i\leq n$ of tiles such that \hbox{$T_0=C$},
\hbox{$T_n=T$} and for all~$i$ in \hbox{$[0..n$$-$$1]$}, \hbox{$T_i\cap T_{i+1}$}
consists of one edge exactly. Then we say that $n$~is the length of the path.
The distance from~$T$ to~$C$ is the shortest length for a path joining~$T$ to~$C$.
Clearly, the distance is always defined. Now, a ball~$B$ around~$C$ of radius~$\rho$
is the set of tiles~$T$ whose distance to~$C$ is at most~$\rho$. The border of~$B$, centered
at~$C$ and denoted by $\partial B$ is the set of tiles whose distance to~$C$ is the radius
of~$B$.  

    Consider~$\Gamma$ a grammar defined in Section~\ref{grammars}. Then, we call 
{\bf $\Gamma$-contour word} the set of words obtained by taking the restriction of a tiling
generated by $\Gamma$ with $Z$ at the central tile~$C$ on the border of a ball of
radius~$n$ around~$C$. As proved in~\cite{mmISRN}, the set of these words is generated
by a 2-iterated pushdown automaton. We reproduce the algorithm which proves this property
in Automaton~\ref{autom2}. It was also mentioned in~\cite{mmISRN} that a simple
application of Ogden's pumping lemma shows that the set of these words cannot be
generated by a pushdown automaton.

\vtop{
\vskip 10pt
\ligne{\hfill
\includegraphics[scale=0.45]{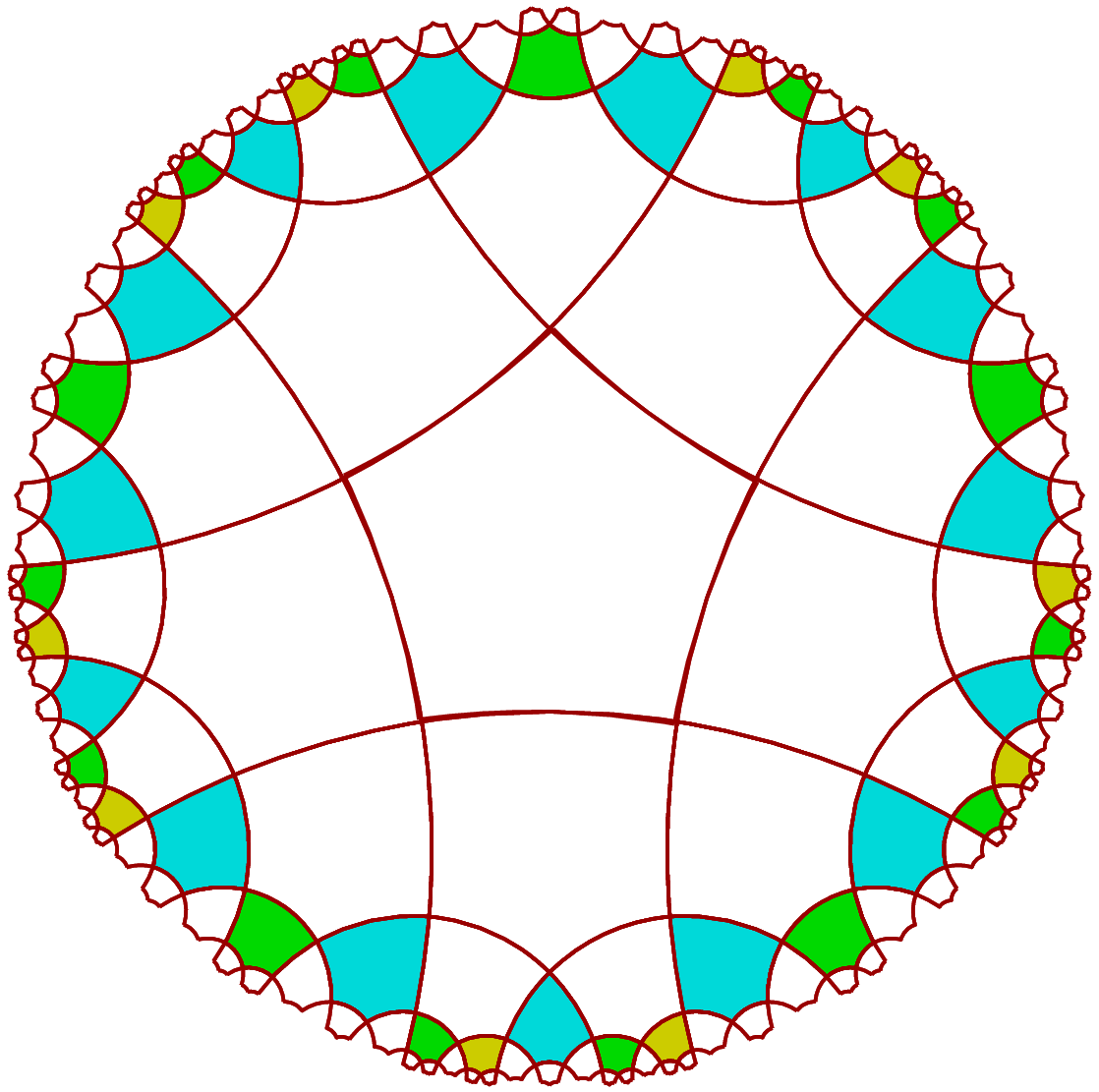}\hfill
\includegraphics[scale=0.45]{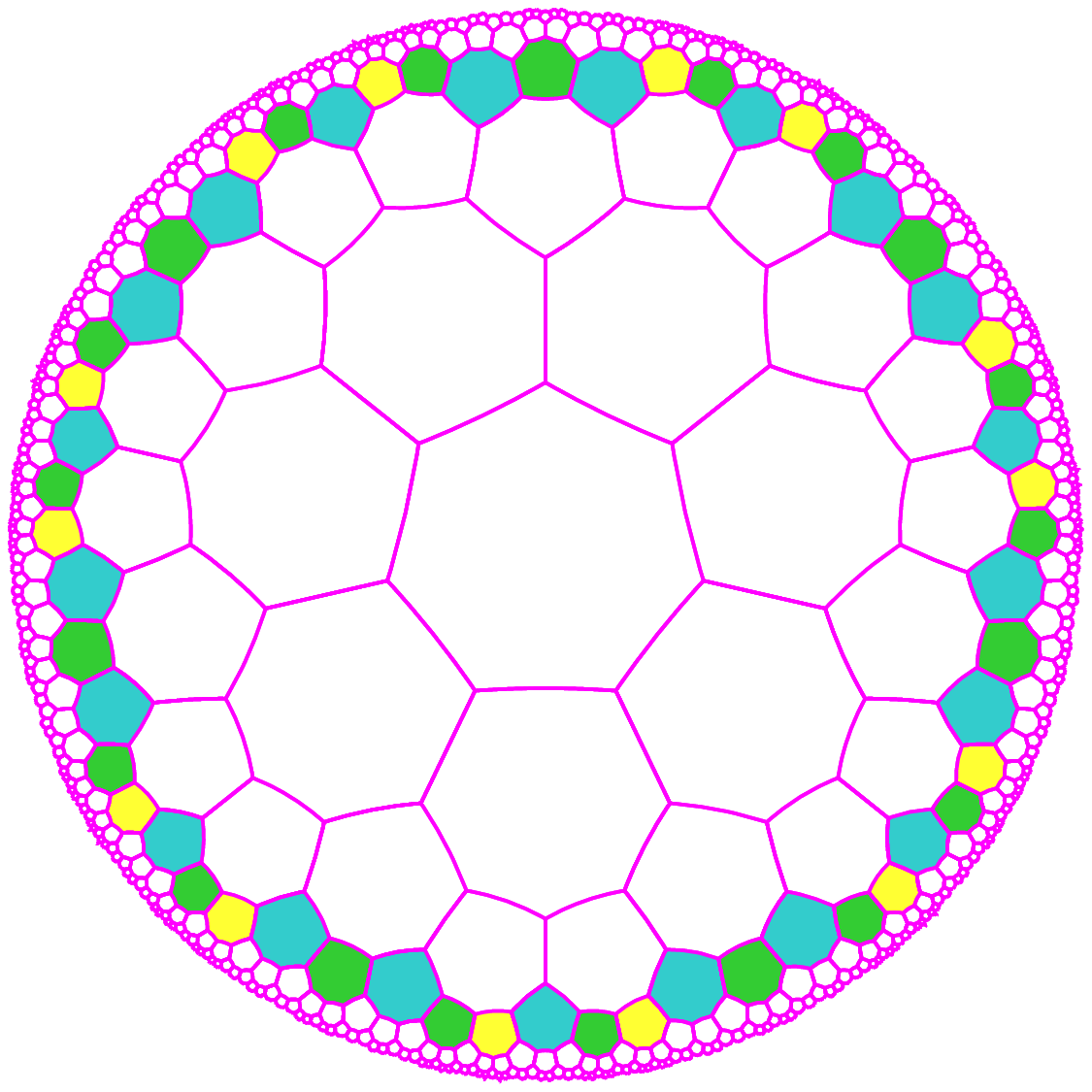}
\hfill
}
\begin{fig}\label{contours}
\leurre
Levels in the heptagrid.
\end{fig}
}

\vtop{
\begin{automaton}\label{autom2}
\leurre
The $2$-pushdown automaton recognizing the contour word of a ball in the pentagrid or
in the heptagrid.
\end{automaton}
\vspace{-12pt}
\grostrait
\vskip 0pt
two states: $q_0$ and~$q_1$; input word in $\{b,w\}^*$;
$\Gamma = \{Z,B,W,F\}$;
\vskip 0pt\noindent
initial state: $q_0$; initial stack: $Z[\epsilon]$; transition function $\delta$:
\vskip 0pt
\vspace{-4pt}
\demitrait
\vskip 0pt
{\leftskip 20pt\parindent 0pt
$\delta(q_0,\epsilon,Z) = \{(q_0,push_2(F)),(q_0,push_1(W^\alpha))\}$\vskip 2pt
$\delta(q_0,\epsilon,ZF) = \{(q_0,push_2(FF)),(q_0,push_1(W^\alpha))\}$\vskip 2pt
$\delta(q_0,\epsilon,WF) = (q_1,pop_2)$\vskip 2pt
$\delta(q_0,\epsilon,BF) = (q_1,pop_2)$\vskip 2pt
$\delta(q_0,b,B) = (q_0,pop_1)$\vskip 2pt
$\delta(q_0,w,W) = (q_0,pop_1)$\vskip 2pt
$\delta(q_1,\epsilon,WF) = (q_0,push_1(BWW))$\vskip 2pt
$\delta(q_1,\epsilon,BF) = (q_0,push_1(BW))$\vskip 2pt
$\delta(q_1,\epsilon,W) = (q_0,push_1(BWW))$\vskip 2pt
$\delta(q_1,\epsilon,B) = (q_0,push_1(BW))$\vskip 2pt
\par}
\vskip 0pt
\demitrait
}
\vskip 10pt

    Now, it was proved in~\cite{mmbook1} that the set of tiles which are on the same
level in a Fibonacci tree belong to a part of the border of a ball around the root
of the Fibonacci tree. Consider again a fixed ball around~$C$ and fix one of the 
finite Fibonacci trees generated around~$C$, say~$\cal F$. We can imagine~$C$ as the central 
tile in Figure~\ref{biinfty_words}. Let $B$ the ball around~$C$ which contains~$\cal F$ and
whose border contains the leaves of~$\cal F$. It is not difficult to find a tile~$C_1$ which 
is a neighbour of~$C$ and such that $C_1$~is the root of a Fibonacci tree ${\cal F}_1$ in the 
ball~$B_1$ around~$C_1$ containing~$\cal F$. We can assume that, in the same way,
the border of~$B_1$ contains the leaves of~${\cal F}_1$. 

In Figure~\ref{biinfty_words}, left-hand side picture, we have a 
line~$\delta_1$ which passes through the mid-points of consecutive edges of heptagons. We 
define~$C_1$ as the yellow neighbour of~$C$ which is cut by~$\delta_1$ and which is 
above~$C$. We can remark that~$C$ is the image of~$C_1$ by a shift 
along the line~$\delta_1$. Now,
it is not difficult to see that the restriction of the tiling to ${\cal F}_1$
contains the restriction of the tiling to~$\cal F$. We can also see that the leaves
of~$\cal F$ are contained in those of~${\cal F}_1$. In the left-and side picture of 
Figure~\ref{biinfty_words}, the sector generated by a black tile is delimited
by the ray~$a$ and the line~$\delta_1$. The two sectors generated by a white tile
are delimited by the line~$\delta_1$ and the ray~$b$ and then by the ray~$b$ and the ray~$e$.
A similar convention is followed for the tree rooted at~$C_1$: the rays~$a_1$,
$b_1$ and~$e_1$ play the same role for~$C_1$ as the rays $a$, $b$ and~$e$ for~$C$.
From the figure, it is not difficult to see that, by induction, we construct
a sequence of tiles~$C_{n}$ with \hbox{$C_0$} a tile crossed by~$\delta_1$ and which is fixed
once and for all, $C_{n+1}$, \hbox{$n\geq0$}, is the neighbour of~$C_n$
which is crossed by~$\delta_1$ and which is defined by the fact that its distance from~$C_0$
is~\hbox{$n$+1} and by the fact that~$C_n$ is in between~$C_{n+1}$ and~$C_0$. We define $B_n$ as 
the ball around~$C_n$ whose border contains~$C_0$ and $F_n$
is the Fibonacci tree rooted at~$C_n$ whose leaves are on~$B_n$. This allows us to
define a sequence of words~$w_n$ which is the trace of the leaves of~$F_n$: 
$w_n$ is in \hbox{$\{B,W\}^\star$} and the $j^{\rm th}$~letter of~$w_n$ is $B$, $W$, depending
on whether the $j^{\rm th}$~leave of~$F_n$ is black, white respectively. The construction shows us
that $w_n$ is a factor of~$w_{n+1}$ and we may assume that
there are nonempty words $u_n$ and~$v_n$ such that \hbox{$w_{n+1}=u_nw_nv_n$}.

   A closer look at the construction indicated in Section~\ref{grammars} shows 
that \hbox{$v_n=w_n$} and that \hbox{$u_{n+1} = u_nw_n$}. Indeed, the separation
between~$u_n$ and~\hbox{$w_nv_n=w_nw_n$} is materialized by~$\delta_1$. Note that
the separation between the two occurrences of~$w_n$ is not fixed: it moves and tends to
infinity as the length of~$w_n$ itself tends to infinity.
And so, $w_n$ is defined at the same time as~$u_n$ by the two equations:
\vskip 10pt
\setbox110=\vtop{\leftskip 0pt\parindent 0pt\hsize=70pt
\ligne{$w_{n+1} = u_nw_nw_n$\hfill}
\ligne{$u_{n+1} = u_nw_n$\hfill}
}
\ligne{\hfill\box110\hfill}
\vskip 10pt
\noindent
with initial conditions \hbox{$u_0 = B$} and \hbox{$w_0 = W$}.

   As the lengths of~$u_n$ and~$w_n$ tend to infinity, and as~$\delta_1$ is fixed, we can 
see from the left-hand side picture of Figure~\ref{biinfty_words} that the sequence of 
words~$w_n$ tend to a bi-infinite word, {\it i.e.} a word whose both ends tend to infinity.
 
\vtop{
\vskip 10pt
\ligne{\hfill
\includegraphics[scale=0.45]{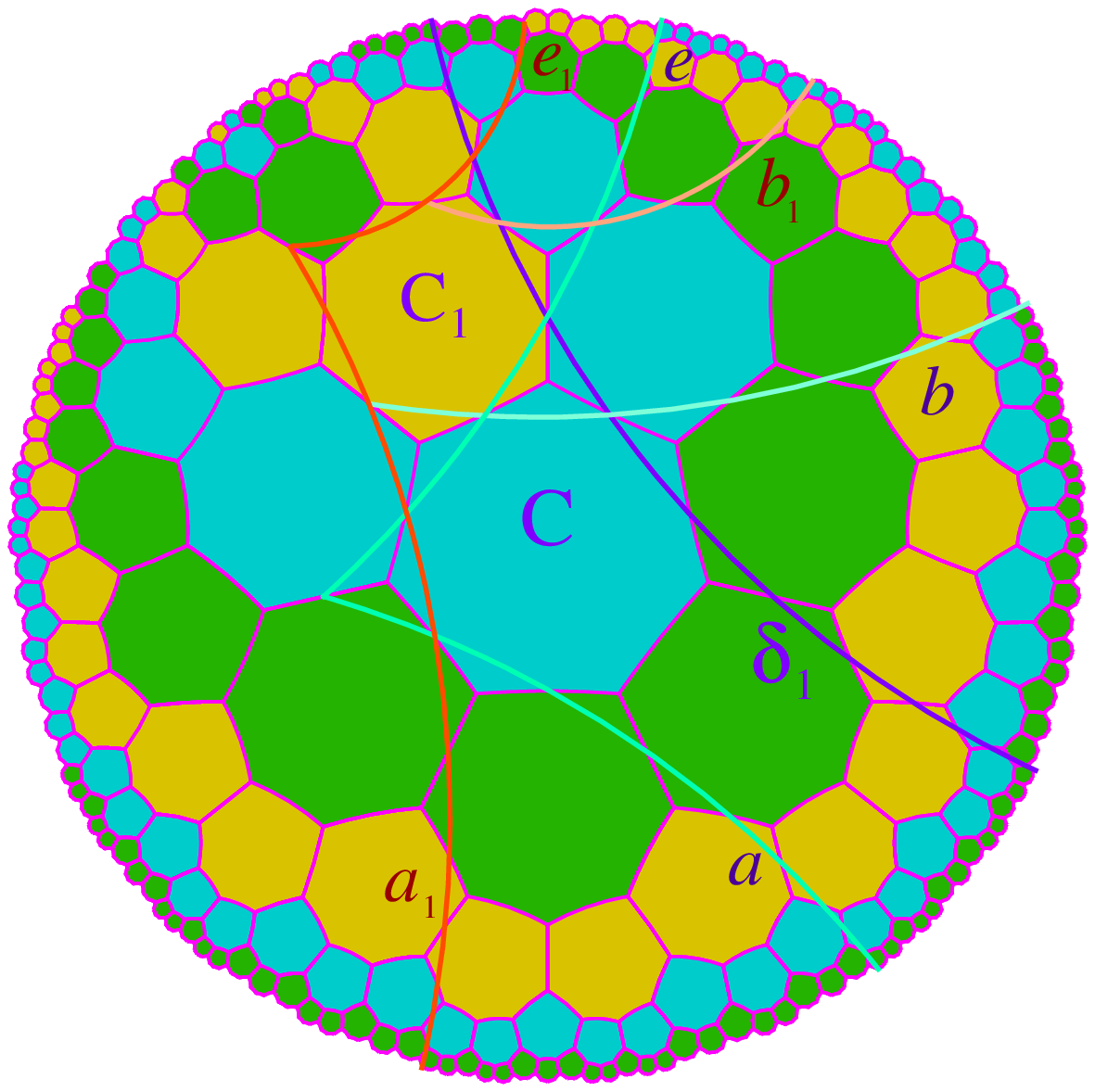}\hfill
\includegraphics[scale=0.45]{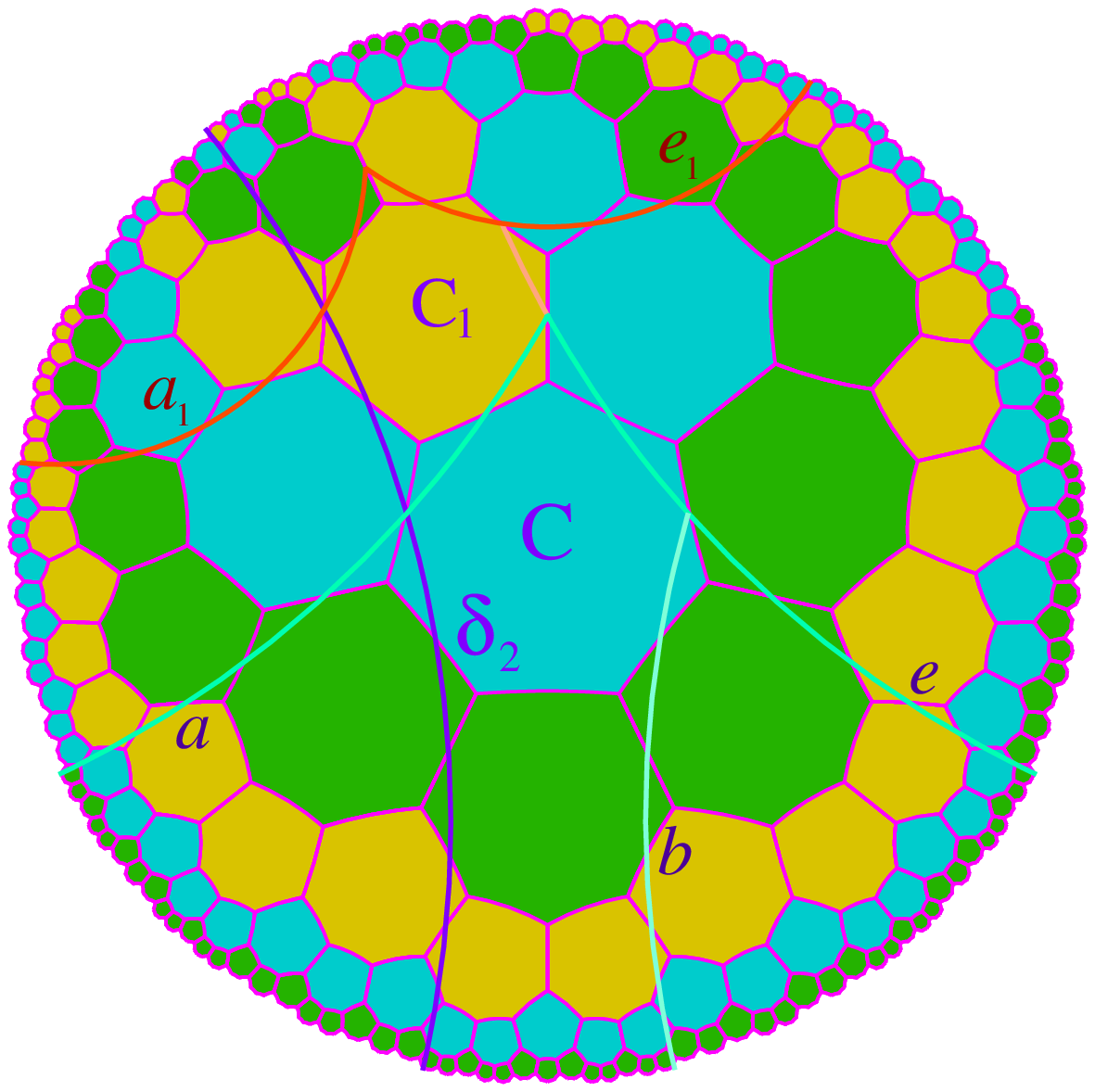}
\hfill
}
\begin{fig}\label{biinfty_words}
\leurre
Heptagrid: construction of bi-infinite words.
\vskip 0pt
Left-hand side: the bi-infinite word associated with the grammar~$(G_1)$.
Right-hand side: the bi-infinite word associated with the grammar~$(G_2)$
\end{fig}
}

   In the right-hand side picture of Figure~\ref{biinfty_words}, we have a similar
construction with the grammar~$(G_2)$. Presently, define $x_n$ and~$y_n$ as the words defined 
by the trace of the leaves of the tree constructed according to the rules of~$(G_2)$ with
\hbox{$x_0=B$} and \hbox{$y_0=W$}. Then, the equations satisfied by
$x_n$ and~$y_n$ are:   
\vskip 10pt
\setbox110=\vtop{\leftskip 0pt\parindent 0pt\hsize=70pt
\ligne{$y_{n+1} = y_nx_ny_n$\hfill}
\ligne{$x_{n+1} = x_ny_n$\hfill}
}
\ligne{\hfill\box110\hfill}
\vskip 10pt
   Note that these words are very different from the~$w_n$'s and the~$u_n$'s.

We can see that, this time, the sectors are delimited
in a different way: the rays~$a$ and~$b$ are not on the same side with respect to~$\delta_1$.
On the figure, we can see that the sectors are delimited  as follows: $a$~and~$\delta_2$ delimit
a white sector, then $\delta_2$ and the ray~$b$ delimit the black sector and, again, we have a white
sector delimited by~$b$ and~$e$. These rays are used for the tree rooted at~$C$. Similar rays,
$a_1$, $b_1$ and~$e_1$ are used for the tree rooted at~$C_1$: as can be seen on the figure,
the tree contains the one defined from~$C$. Note that $b_1$ is the continuation of~$e$.
As in the case with the left-hand side picture,
this picture also defines a bi-infinite word as the limit of~$w_n$.

   Note that, in both case, $u_n$~tends to a limit which is infinite on one side only: this can be
seen by the fact that the black sector is always delimited by~$\delta_1$ or~$\delta_2$ and these
lines are fixed. The infinite limit is finite to the left in the case of~$(G_1)$, it is
finite to the right in the case of~$(G_2)$. 

\vtop{
\vskip 10pt
\ligne{\hfill
\includegraphics[scale=0.45]{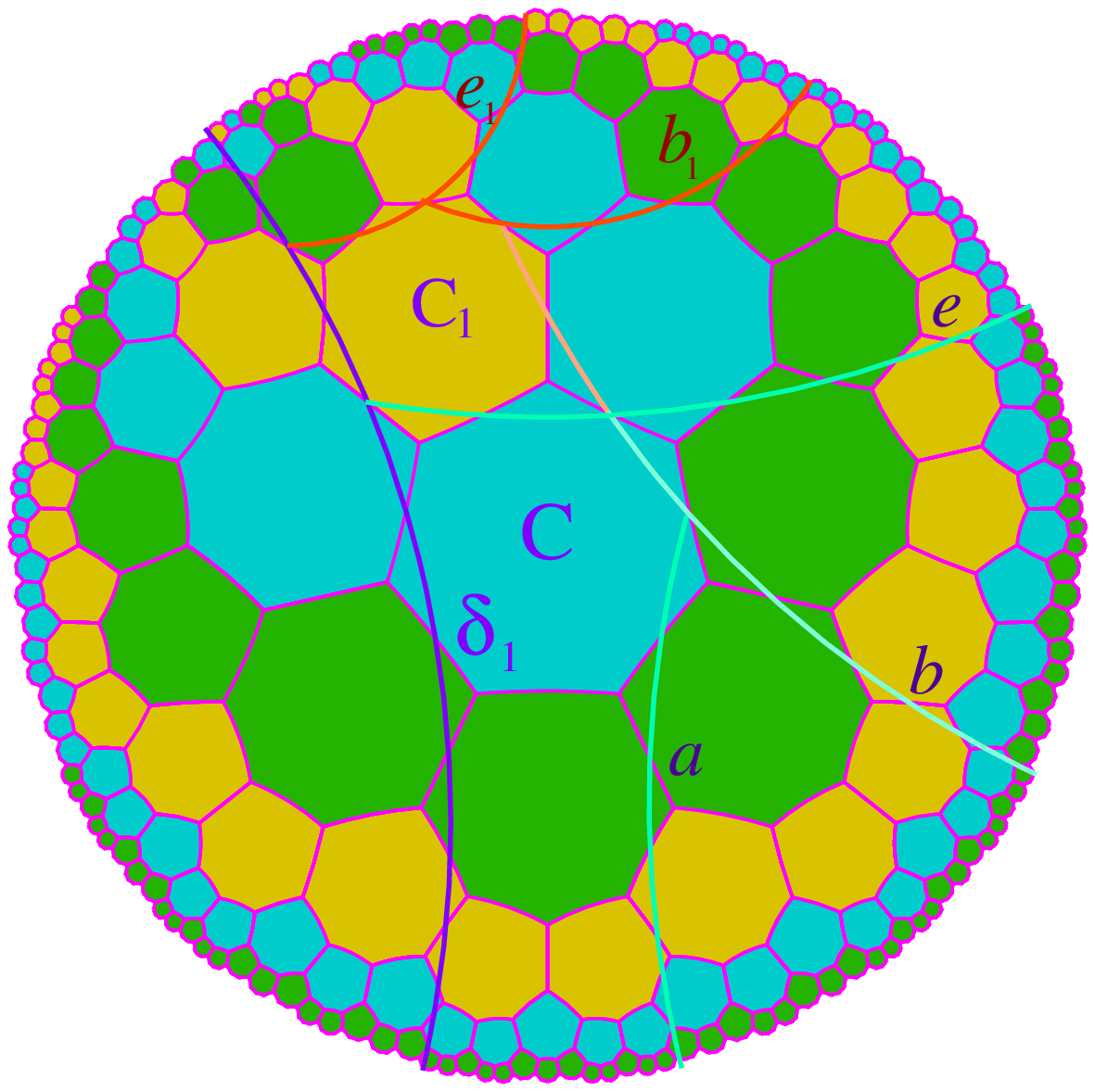}\hfill
\includegraphics[scale=0.45]{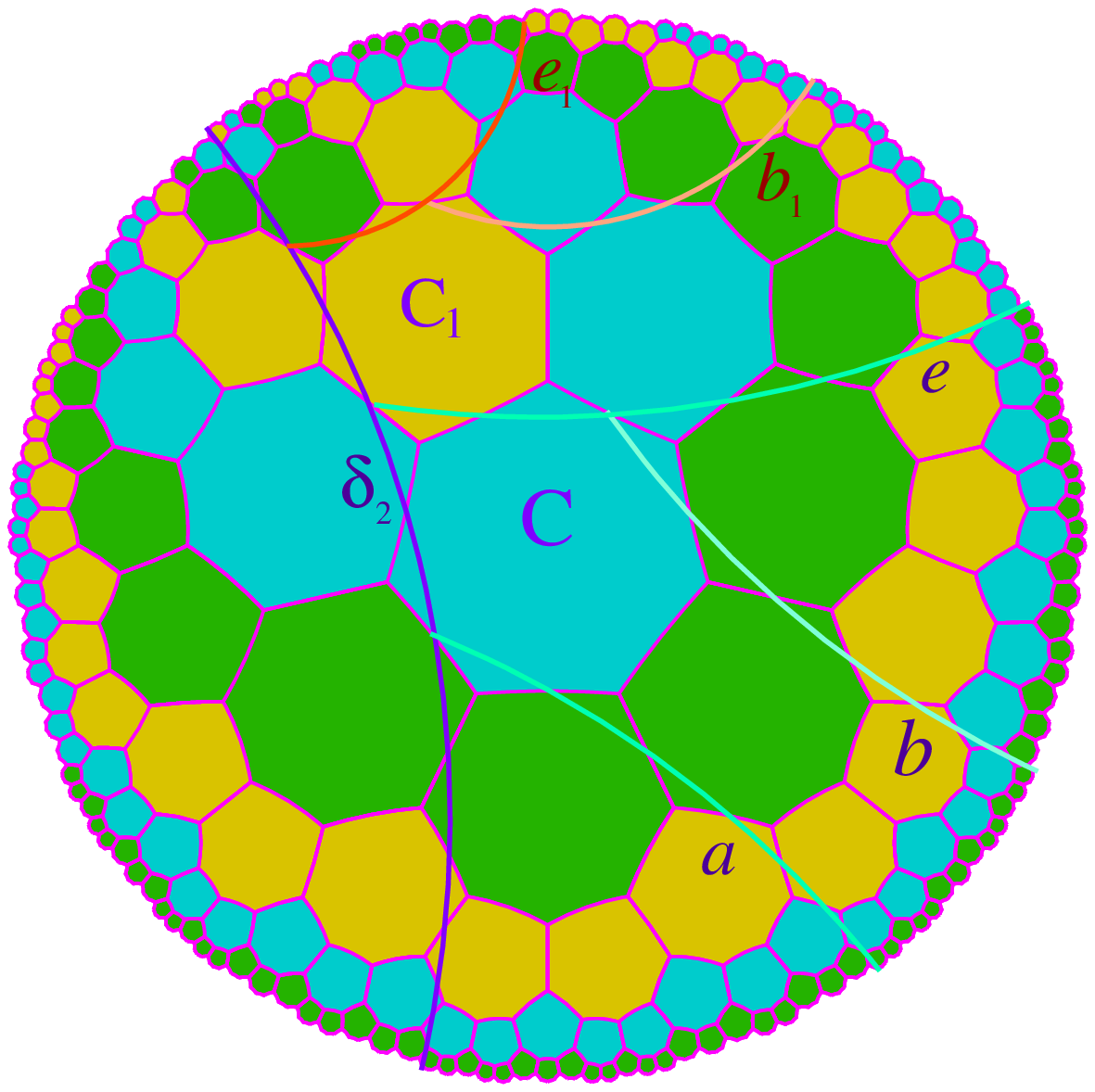}
\hfill
}
\begin{fig}\label{mono_infty_words}
\leurre
Heptagrid: construction of one-sided infinite words.
\vskip 0pt
Left-hand side: an infinite word associated with the grammar~$(G_1)$,
Right-hand side: an infinite word associated with the grammar~$(G_2)$
\end{fig}
}


      In both cases, say that $\delta_1$ and~$\delta_2$ are {\bf separators}: $\delta_1$
separates $u_n$ from~$w_nw_n$ for each~$n$; $\delta_2$ separates $y_n$ from~$x_ny_n$.

    Figure~\ref{mono_infty_words} illustrates a similar construction leading to
an infinite limit for~$w_n$ which is infinite on one side only. The rays $a$, $b$ and~$e$
play similar roles with the lines~$\delta_1$ or~$\delta_2$ as in the Figure~\ref{biinfty_words}.
Note that in the left-hand side picture, $a_1$ is not mentioned as it contains the ray~$b$.
In the right-hand side picture, the ray~$a_1$ coincide with the ray~$e$. From the
picture, it is clear that this time the limit of~$w_n$ and that of~$y_n$ are both infinite to 
the right. In the case of~$(G_1)$, the limit of~$u_n$ is also infinite to the right only.
In the right-hand side picture, we can see that the different terms~$x_n$ are disjoint.
However, each one is the same asi~$u_n$ with the same index in the left-hand side
picture: accordingly, the limit is the same. 

   Other constructions of the same type, with again a fixed separator between both 
occurrences of~$w_n$ in the case of~$G_1$ and in between~$u_n$ and~$w_n$ in the second case lead 
to different pictures and to other infinite words. We leave them as an exercise to the reader.

\subsection*{Acknowledgment}

The second author acknowledges support by the project Universit\'e de la Grande R\'egion
UniGR that enabled his visit during February 2013, in particular at LORIA, Universit\'e 
de Lorraine, France.

\nocite{*}
\bibliographystyle{eptcs}
\bibliography{mmkgs_mcu_final_12}

\begin{thebibliography}{10}
\providecommand{\bibitemdeclare}[2]{}
\providecommand{\surnamestart}{}
\providecommand{\surnameend}{}
\providecommand{\urlprefix}{Available at }
\providecommand{\url}[1]{\texttt{#1}}
\providecommand{\href}[2]{\texttt{#2}}
\providecommand{\urlalt}[2]{\href{#1}{#2}}
\providecommand{\doi}[1]{doi:\urlalt{http://dx.doi.org/#1}{#1}}
\providecommand{\bibinfo}[2]{#2}

\bibitemdeclare{article}{fratani-senizergues}
\bibitem{fratani-senizergues}
\bibinfo{author}{S.~Fratani \surnamestart G.~S\'enizergues\surnameend}
  (\bibinfo{year}{2006}): \emph{\bibinfo{title}{Iterated pushdown automata and
  sequences of rational numbers}}.
\newblock {\sl \bibinfo{journal}{Annals of pure and applied logic}}
  \bibinfo{volume}{141}, pp. \bibinfo{pages}{363--411},
  \doi{10.1016/j.apal.2005.12.004}.

\bibitemdeclare{article}{mmgsFI}
\bibitem{mmgsFI}
\bibinfo{author}{M.~Margenstern \surnamestart G.~Skordev\surnameend}
  (\bibinfo{year}{2003}): \emph{\bibinfo{title}{Tools for devising cellular
  automata in the hyperbolic 3D space}}.
\newblock {\sl \bibinfo{journal}{Fundamenta Informaticae}}
  \bibinfo{volume}{58}(\bibinfo{number}{2}), pp. \bibinfo{pages}{369--398}.

\bibitemdeclare{article}{gms-alg}
\bibitem{gms-alg}
\bibinfo{author}{C.~\surnamestart Goodman-Strauss\surnameend}
  (\bibinfo{year}{2009}): \emph{\bibinfo{title}{Regular production systems and
  triangle tilings}}.
\newblock {\sl \bibinfo{journal}{Theoretical Computer Science}}
  \bibinfo{volume}{410}, pp. \bibinfo{pages}{1534--1549},
  \doi{10.1016/j.tcs.2008.12.012}.

\bibitemdeclare{article}{greibach}
\bibitem{greibach}
\bibinfo{author}{S.~\surnamestart Greibach\surnameend} (\bibinfo{year}{1970}):
  \emph{\bibinfo{title}{Full AFL's and nested iterated substitution}}.
\newblock {\sl \bibinfo{journal}{Information and Control}}
  \bibinfo{volume}{16}(\bibinfo{number}{1}), pp. \bibinfo{pages}{7--35},
  \doi{10.1016/s0019-9958(70)80039-0}.

\bibitemdeclare{book}{kelarev}
\bibitem{kelarev}
\bibinfo{author}{A.V. \surnamestart Kelarev\surnameend} (\bibinfo{year}{2003}):
  \emph{\bibinfo{title}{Graph Algebras and Automata}}.
\newblock \bibinfo{publisher}{Marcel Dekker}, \bibinfo{address}{New York}.

\bibitemdeclare{article}{mmJUCSii}
\bibitem{mmJUCSii}
\bibinfo{author}{M.~\surnamestart Margenstern\surnameend}
  (\bibinfo{year}{2000}): \emph{\bibinfo{title}{New Tools for Cellular Automata
  of the Hyperbolic Plane}}.
\newblock {\sl \bibinfo{journal}{Journal of Universal Computer Science}}
  \bibinfo{volume}{6}(\bibinfo{number}{12}), pp. \bibinfo{pages}{1226--1252},
  \doi{10.3217/jucs-006-12-1226}.

\bibitemdeclare{article}{mmJUCStile}
\bibitem{mmJUCStile}
\bibinfo{author}{M.~\surnamestart Margenstern\surnameend}
  (\bibinfo{year}{2002}): \emph{\bibinfo{title}{Tiling the hyperbolic plane
  with a single pentagonal tile}}.
\newblock {\sl \bibinfo{journal}{Journal of Universal Computer Science}}
  \bibinfo{volume}{8}(\bibinfo{number}{2}), pp. \bibinfo{pages}{297--316},
  \doi{10.3217/jucs-008-02-0297}.

\bibitemdeclare{article}{mmJUCS4D}
\bibitem{mmJUCS4D}
\bibinfo{author}{M.~\surnamestart Margenstern\surnameend}
  (\bibinfo{year}{2004}): \emph{\bibinfo{title}{The tiling of the hyperbolic
  $4D$ space by the 120-cell is combinatoric}}.
\newblock {\sl \bibinfo{journal}{Journal of Universal Computer Science}}
  \bibinfo{volume}{10}(\bibinfo{number}{9}), pp. \bibinfo{pages}{1212--1238},
  \doi{10.3217/jucs-010-09-1212}.

\bibitemdeclare{book}{mmbook1}
\bibitem{mmbook1}
\bibinfo{author}{M.~\surnamestart Margenstern\surnameend}
  (\bibinfo{year}{2007}): \emph{\bibinfo{title}{Cellular Automata in Hyperbolic
  Spaces, volume I : Theory}}, \bibinfo{edition}{first} edition.
\newblock {\sl \bibinfo{series}{Advances in Unconventional Computing and
  Cellular Automata, Editor: Andrew Adamatzky}}~\bibinfo{volume}{1},
  \bibinfo{publisher}{Old City Publishing-\'Edition des archives
  contemporaines}, \bibinfo{address}{Philadelphia, PA, USA - Paris, France}.

\bibitemdeclare{book}{mmbook2}
\bibitem{mmbook2}
\bibinfo{author}{M.~\surnamestart Margenstern\surnameend}
  (\bibinfo{year}{2008}): \emph{\bibinfo{title}{Cellular Automata in Hyperbolic
  Spaces, volume II : Implementation and Computations}},
  \bibinfo{edition}{first} edition.
\newblock {\sl \bibinfo{series}{Advances in Unconventional Computing and
  Cellular Automata, Editor: Andrew Adamatzky}}~\bibinfo{volume}{2},
  \bibinfo{publisher}{Old City Publishing-\'Edition des archives
  contemporaines}, \bibinfo{address}{Philadelphia, PA, USA - Paris, France}.

\bibitemdeclare{article}{mmISRN}
\bibitem{mmISRN}
\bibinfo{author}{M.~\surnamestart Margenstern\surnameend}
  (\bibinfo{year}{2012}): \emph{\bibinfo{title}{An Application of Iterative
  Pushdown Automata to Contour Words of Balls and Truncated Balls in Hyperbolic
  Tessellations}}.
\newblock {\sl \bibinfo{journal}{ISRN Algebra}} \bibinfo{volume}{2012}, p.
  \bibinfo{pages}{14p}, \doi{10.5402/2012/742310}.

\bibitemdeclare{article}{maslov}
\bibitem{maslov}
\bibinfo{author}{A.N. \surnamestart Maslov\surnameend} (\bibinfo{year}{1974}):
  \emph{\bibinfo{title}{The hierarchy of indexed languages}}.
\newblock {\sl \bibinfo{journal}{Soviet Mathematics, Doklady}}
  \bibinfo{volume}{15}, pp. \bibinfo{pages}{1170--1174}.

\end{thebibliography}

\end{document}